\documentclass{ws-spin}
\usepackage{hyperref}
\usepackage{multicol,overcite}
\def\bib{B\kern-.05em{I}\kern-.025em{B}\kern-.08em}
\def\btex{B\kern-.05em{I}\kern-.025em{B}\kern-.08em\TeX}
\begin{document}


\markboth{S. Ghosh, A. Manchon}{Signature of topological phases in Zitterbewegung}

\title{Signature of topological phases in Zitterbewegung}

\author{S. Ghosh$^\dagger$, A. Manchon$^\ddagger$}

\address{Physical Science and Engineering Division (PSE), King Abdullah University of Science and Technology (KAUST), Thuwal 23955, Saudi Arabia\\
\email{$^\dagger$sumit.ghosh@kaust.edu.sa,$^\ddagger$aurelien.manchon@kaust.edu.sa}}

\maketitle


\begin{abstract}
We have studied the {\em Zitterbewegung} effect on an infinite two dimensional sheet with honeycomb lattice. By tuning the perpendicular electric field and the magnetization of the sheet, it can enter different topological phases. We have shown that the phase and magnitude of Zitterbewegung effect, i.e. the jittering motion of relativistic particles, correlates with the various topological phases. The topological phase diagram can be reconstructed by analyzing these features. Our findings are applicable to materials like silicene, germanene, stanene etc. 
\end{abstract}


\begin{multicols}{2}

\section{Introduction}
The rise of graphene and its two dimensional siblings (silicene, germanene, stanene etc.) has significantly lowered the barrier between high and low energy physics. The Dirac fermion-like behavior of low energy electrons in these materials offers an appealing platform to investigate the predictions from high-energy physics in condensed matter \cite{Semenoff1984, Geim2007, Young2008, Romanovsky2013}. A remarkable aspect of two-dimensional hexagonal lattices is that they exhibit topological phase transitions. By tuning the material's parameters, transitions from trivial (metallic or insulating) phase to quantum spin/anomalous/valley Hall phases can be achieved \cite{Ezawa2012,Han2014}. These non-trivial topological phases are usually determined either by calculating the Berry phase or Chen number of the bulk material or by probing the existence of quantized edge states. 
The reason why two-dimensional hexagonal lattices display such a variety of phases is the presence of Dirac kinetic term that couples the carrier momentum with its (pseudo)spin degree of freedom, together with the emergence of an orbital gap \cite{Pesin2012,Han2014}. One of the outcomes of this (pseudo)spin-momentum locking is the {\em Zitterbewegung} effect \cite{Lock1979}, which arises from the interference between positive and negative energy states and causes an oscillatory motion for relativistic free particle. The effect, originally proposed in the context of relativistic particles, has also been predicted to occur in condensed matter \cite{Schliemann2005, Zawadzki2005, Cserti2006, Zawadzki2011}  and verified in several experiments with photonic crystal, cold atoms, trapped ions, Bose-Einstein condensate as well \cite{Zhang2008, Vaishnav2008, Gerritsma2010, LeBlanc2013, Qu2013}.

Although this effect has been extensively studied in two-dimensional materials \cite{Katsnelson2006a}, previous investigations mainly focused on the occurrence of Zitterbewegung and its behavior under a magnetic field \cite{Rusin2007, Yudin2014, Romera2014}. In practice Zitterbewegung is a very rich phenomena that provides a lot of information about the system. For example, it is associated with the origin of spin in a relativistic system \cite{Barut1981} and can be exploited to control the spin polarized orbital motion of electrons \cite{Shi2013}. Furthermore, while previous studies have all focused on the $transverse$ Zitterbewegung, a lot of interesting features are hidden in the $longitudinal$ Zitterbewegung as well. Recently it has been shown that Zitterbewegung can manifest the presence of an edge state in zigzag graphene nanoribbon by the emergence of a resonance \cite{Ghosh2015}. This brings us to an obvious question: does Zitterbewegung also carry information about the topological phases of the system? It has been shown that the Zitterbewegung amplitude shares a close connection with Berry curvature and Chern number in a multiband system \cite{David2010, Cserti2010}. However it is not clear how the oscillations behave in different topological phases.

In this paper we demonstrate that Zitterbewegung can be used to probe the different topological phases of two-dimensional hexagonal lattices. For our study we choose silicene, a two dimensional buckled
honeycomb lattice with strong spin-orbit coupling, where one can tune the topological phases with an external electric field and onsite magnetization \cite{Ezawa2012}. We present a systematic analysis of Zitterbewegung in different topological phases that enable us to recover the topological phase diagram of the material. The formalism we adopted here is quite generic and hence applicable to any Dirac material.

\section{Wavepacket Evolution}
The Hamiltonian for silicene (germanene or stanene) near $K$ and $K'$ points is given by \cite{Ezawa2012}

\begin{eqnarray}
\hat{H}_\eta &=& \hbar v_F (\eta k_x \hat{\tau}_x + k_y \hat{\tau}_y ) + \eta \hat\tau_z \hat{h}_{11} - \ell E_Z \hat{\tau}_z + M \hat{\sigma}_z  \nonumber \\
 && + \lambda_{R1} (\eta \hat{\tau}_x \hat{\sigma}_y - \hat{\tau}_y \hat{\sigma}_x)/2, \nonumber \\
\hat{h}_{11} &=& \lambda_{SO} \hat{\sigma}_z + a \lambda_{R2} (k_y \hat{\sigma}_x - k_x \hat{\sigma}_y),
\label{h1}
\end{eqnarray}
where $\eta = \pm 1$ corresponds $K$ and $K'$ valley, $\hat{\bm\sigma}$ and $\hat{\bm\tau}$ are Pauli matrices for spin and valley, $a$ is the interatomic distance and $\ell$ is the buckling height. $E_Z$ is an external field applied perpendicular to the plane and $M$ is the onsite magnetization. $\lambda_{SO}$ is the spin-orbit coupling, $\lambda_{R}$ is the second nearest Rashba parameter. The system undergoes a topological phase transition at a critical electric field $E_C = \lambda_{SO}/\ell$ for $M=0$. The values of the parameters for different materials are given in Table \ref{param}.

\begin{tablehere}
\tbl{Different parameter values for graphene (Gr), silicene (Si), germanene (Ge) and stanene (Sn) \cite{Liu2011,Ezawa2015a}. }
{\begin{tabular}{@{}|c|c|c|c|c|c|c|@{}}
\toprule
Atom & $a$ & $\ell $ & $\lambda_{SO} $ & $ \lambda_{R}$ & $v_F$ & $E_C$ \\
 & $\rm (\AA)$ & $\rm (\AA)$ & $\rm (meV)$ & $ \rm (meV)$  & $\rm 10^5 m/s$ & $ \rm (meV/\AA)$ \\ \colrule
Gr & 2.46 & 0.00 &$10^{-3}$& 0.0 & 9.8 & $\infty$   \\ \colrule
Si & 3.86 & 0.23 &  3.9  & 0.7 & 5.5 & 17  \\ \colrule
Ge & 4.02 & 0.33 & 43.0  &10.7 & 4.6 &130.3 \\ \colrule 
Sn & 4.70 & 0.40 & 100   & 9.5 & 4.9 &250   \\ \botrule 
\end{tabular}}
\label{param}
\end{tablehere}

In practice $\lambda_{R}\ll \lambda_{SO}$ and we can drop this term. In that case the spin up and down Hamiltonians decouple and we can write them as $2 \times 2$ matrices,

\begin{eqnarray}
H_{\eta,s} = \begin{pmatrix}
- \ell E_Z + s(M + \eta \lambda_{SO}) & v_F (-i k_y + \eta k_x) \\
v_F (i k_y + \eta k_x) & \ell E_Z + s(M-\eta \lambda_{SO})
\end{pmatrix}.\nonumber\\
\label{h2}
\end{eqnarray} 

This Hamiltonian is analytically solvable, and the eigenvalues and eigenfunctions read
\begin{eqnarray}
E_{\eta,s}^\pm &=& \frac{1}{2} ( m_{\eta,s}^A + m_{\eta,s}^B ) \nonumber \\
&& \pm \frac{1}{2} \sqrt{(m_{\eta,s}^A - m_{\eta,s}^B)^2 + 4 v_F (k_x^2+k_y^2)}, \\
\psi_{\eta,s}^\pm(\vec{k}) &=& \begin{pmatrix}
\cos (\Theta_{\eta,s}^\pm) \\ \sin (\Theta_{\eta,s}^\pm)
\end{pmatrix},
\label{epsi}
\end{eqnarray} 
where, $m_{\eta,s}^{A/B} = \mp \ell E_z + s(M+\eta \lambda_{SO})$, $
\Theta_{\eta,s}^\pm = \tan^{-1}\frac{v_F(i k_y + \eta k_x)}{E_{\eta,s}^\pm - m_{\eta,s}^B}$.
Due to the valley dependence of spin-orbit coupling, for $E_Z=\pm E_C$ carriers with opposite spin projection form Dirac cones at opposite valleys (Fig. \ref{band}). Consequently, at a particular valley carriers with opposite spin projection undergo topological transition at a different critical electric field. In other words, as illustrated in Fig. \ref{band}, at K (K') point, spin up (down) band exhibits a Dirac cone at $E_Z=E_C$, while spin down (up) band presents an orbital gap. The situation is reversed for $E_Z=-E_C$.

\begin{figurehere}
\centerline{
\includegraphics[width=8cm]{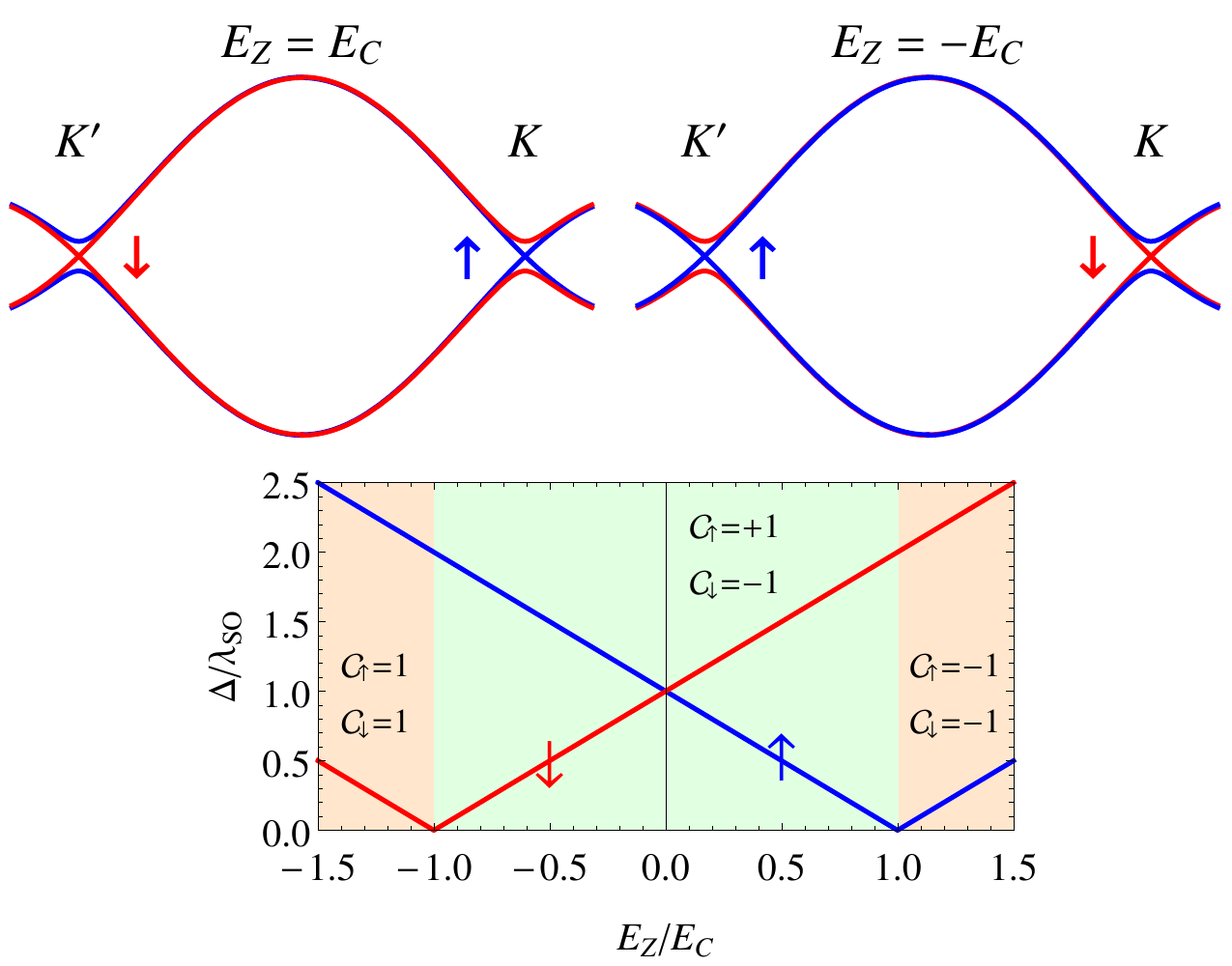}
}
\caption{Top: Dirac cone for $E_Z = \pm E_C$ at $K$ and $K'$ valley. Bottom: Variation of band gap for spin up and down at $K$ valley for $M=0$. The green and orange regions correspond to quantum spin hall phase ($\mathcal{C}_s=1$) and bulk insulator ($\mathcal{C}_s=0$) phase where $\mathcal{C}_s=(\mathcal{C}_\uparrow-\mathcal{C}_\downarrow)/2$ is the spin Chern number.}
\label{band}
\end{figurehere}

Since there is no position dependent term in the Hamiltonian, momentum is a conserved quantity and we can use the momentum eigenstates to create a wave packet. To do so, we use a Gaussian envelop for the momentum distribution so that the wavepacket in the real space is also Gaussian. By choosing a narrow width of the momentum distribution, we can avoid valley mixing. From now on we will focus on $K$ ($\eta=1$) valley only and drop the valley index. The wavepacket can be expressed as \cite{Rusin2007}

\begin{eqnarray}
\Psi_{s}(\vec{r},t) &=& \frac{1}{\mathcal{N}} \int g(\vec{k},\vec{k_0}) \frac{1}{\sqrt{2}} [\psi_s^+(\vec{k})e^{-i E_s^+ t} \nonumber \\
&& + \psi_s^-(\vec{k})e^{-i E_s^- t} ] e^{i \vec{k}.\vec{r}}  d\vec{k}, \\
g(\vec{k},\vec{k_0}) &=& \frac{d}{\sqrt{\pi}} e^{-1/2(\vec{k}-\vec{k_0})^2 d^2},
\label{gpack}
\end{eqnarray}
where $d$ is the width of momentum distribution and $\mathcal{N}$ is the normalization factor. Eq. \ref{gpack} applies to a two band system, but can be generalized to multiband system by adding up all contributing states. Due to the Gaussian envelop, only a selected portion of the bands contributes to the wavepacket as shown in Fig.~\ref{fig1}. If the Fermi level lies in the middle of two bands, then both bands contribute to the wavepacket (Fig.~\ref{fig1}a). On the other hand if the Fermi level cuts one of the bands, only finite region of the selected portion can make a contribution (Fig.~\ref{fig1}b). 

\begin{figurehere}
\centerline{
\includegraphics[width=8cm]{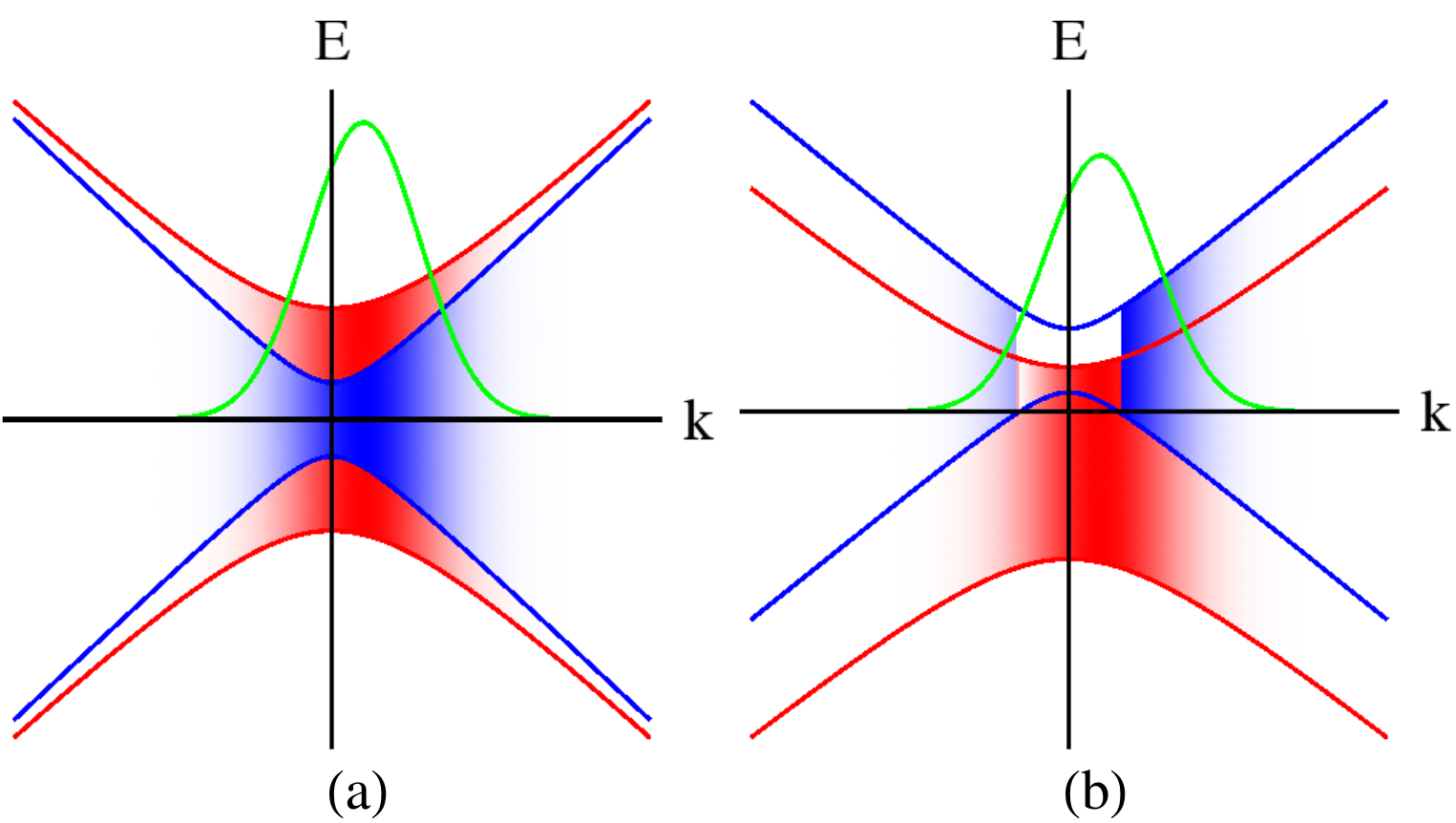}
}
\caption{Formation of Gaussian wavepacket. The blue and red lines corresponds up and down spin bands and the green line shows the Gaussian distribution. The shaded region shows the portion of bands that contributes to the wavepacket.}
\label{fig1}
\end{figurehere}

Once we construct the wavepacket, we can easily calculate the Zitterbewegung component of the position as described in Ref. \cite{Ghosh2015}. Let $\Psi_s^+(\vec{r},t)$ and $\Psi_s^-(\vec{r},t)$ be two wavepackets with the same momentum distribution and made of only positive and negative energy states. The Zitterbewegung component of an operator $\hat{\mathcal{O}}$ is given by
\begin{eqnarray}
\mathcal{O}_{ZB} (s,t) &=& \langle \Psi_s |\hat{\mathcal{O}}| \Psi_s \rangle \nonumber \\
&& - \frac{1}{2}(\langle \Psi_s^+ |\hat{\mathcal{O}}| \Psi_s^+ \rangle + \langle \Psi_s^- |\hat{\mathcal{O}}| \Psi_s^- \rangle),
\label{z}
\end{eqnarray}
where the subscript ZB denotes the Zitterbewegung component. Note that our definition is same as \cite{Rushin2007} where they define  Zitterbewegung component as the sum over expectation values due to overlap integral between different states.

As mentioned above, depending on the values of $M$ and $E_Z$ silicene exhibits different topological phases, which have been described, for instance, in Ref. \cite{Ezawa2012}. The topological phase diagram of silicene when varying $E_Z$ and $M$ is reported in Fig. \ref{xplus} by the dashed lines. In the following, we explore the nature of Zitterbewegung throughout the phase diagram and establish a correlation between the different topological phases and Zitterbewegung features.
To do so, we construct wavepackets in each of the regions of the topological phase diagram of silicene, and calculate the Zitterbewegung component of the coordinates. We choose the central momentum of the wavepacket to be $\vec{k_0} = 0.00025/a \hat{x}$ and the width of the Gaussian distribution to be $d=5000 a$.

First let us focus on two points on the $M=0$ axis, say $E_Z = 1.5E_C$ and $E_Z = 0.5E_C$. The first case corresponds to a topologically trivial phase for both spin up and spin down bands. In the second case, spin up is in topologically nontrivial phase while spin down is in topologically trivial phase.
The band structures for these two cases are reported on the top and bottom left panels of Fig. \ref{e12}, respectively. We also compute the Zitterbewegung contributions to the position of the wavepacket ($x_{ZB}$, $y_{ZB}$) as a function of time. These results are reported on the top and bottom right panels of Fig. \ref{e12} for the $E_Z=0.5E_C$ and $E_Z=1.5E_C$, respectively.

\begin{figurehere}
\centerline{
\includegraphics[scale=0.55]{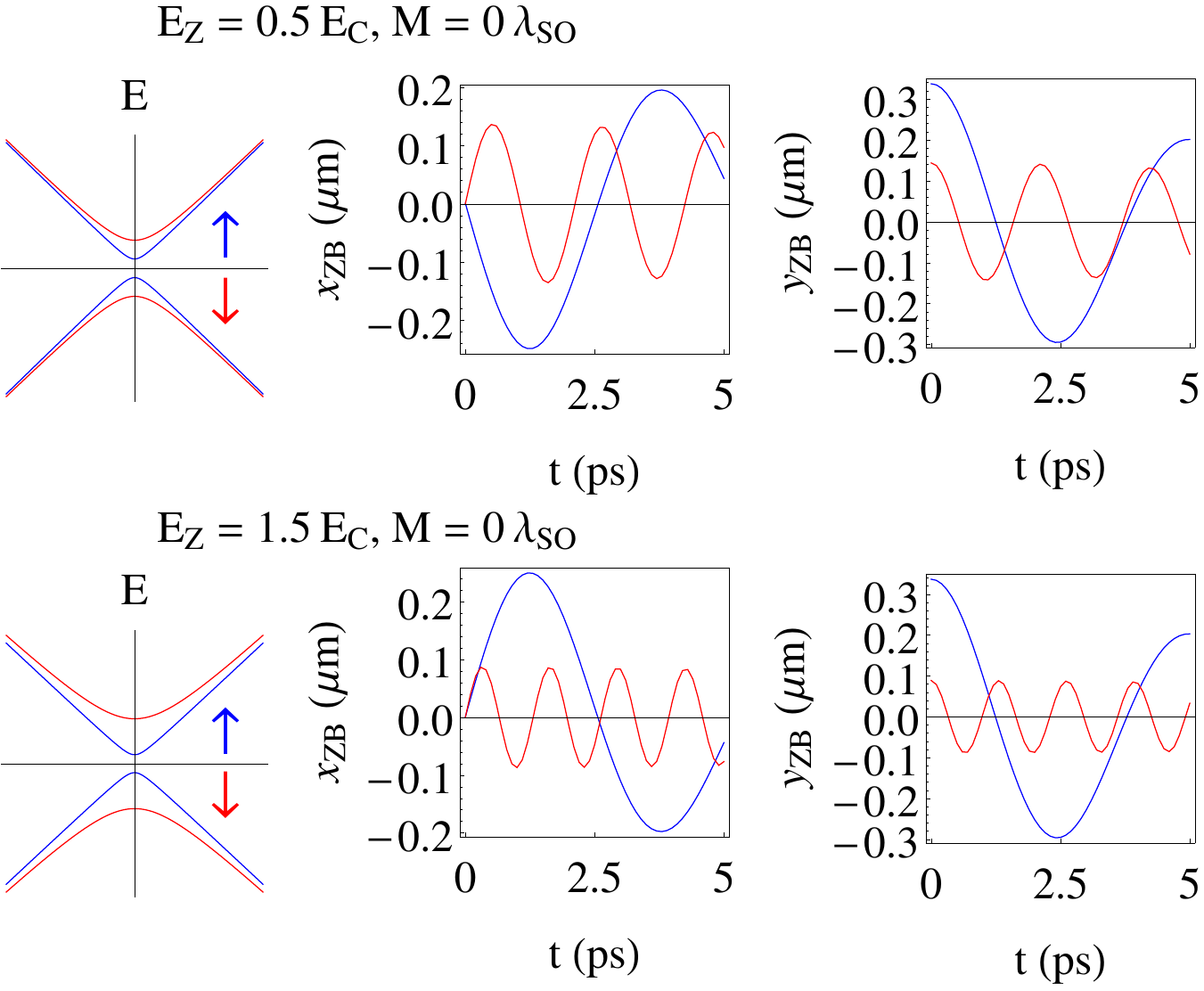}
}
\caption{Zitterbewegung component of $x$ and $y$ coordinates for different values of $E_Z$ and $M$. The left panels show the corresponding band structure.}
\label{e12}
\end{figurehere}

One can readily see that the qualitative nature of the band structures is the same in both cases - up spin having a smaller gap and the down spin having a larger gap. $x_{ZB}$ however behaves completely differently in these two cases. We can see that $x_{ZB}$ for spin up states undergoes a $\pi$ phase shift when moving from topologically nontrivial to trivial phase while the phase for down spin remains the same. $y_Z$ on the other hand does not show any qualitative change. 

Let us focus on $x_{ZB}$ for a spin unpolarized wavepacket.
The change of phase in $x_Z$ oscillation is simply related to the inversion of the band gap through the topological transition.  Depending on the values of $E_{ZB}$ and $M$, different spin components oscillate with different amplitude and frequency resulting in beating in charge and spin density waves, as illustrated on Fig.~\ref{beat}.
\begin{figurehere}
\centerline{
\includegraphics[scale=1]{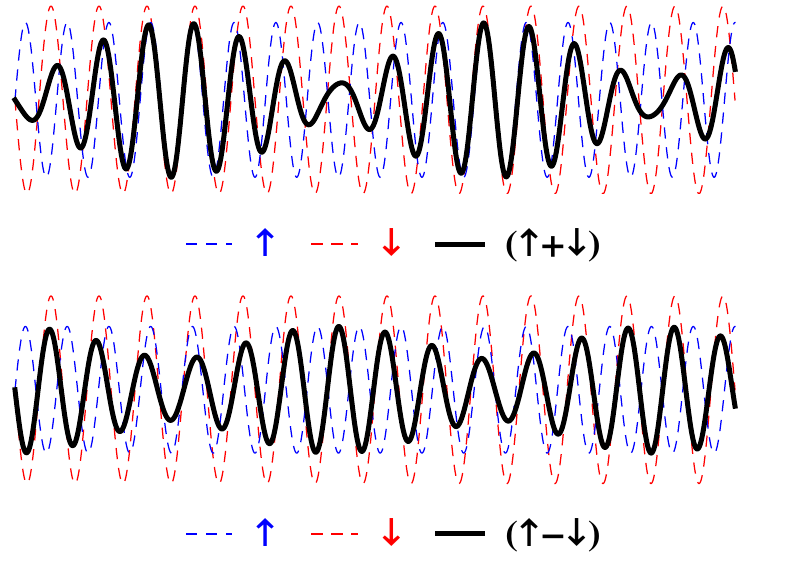}
}
\caption{Formation of beating in charge (top) and spin (bottom) density waves.}
\label{beat}
\end{figurehere}

Modeling the oscillations for individual spins as damped sinusoids ($A e^{-\Gamma t} \sin(\omega t + \delta_0)$) \cite{Ghosh2015} we can easily evaluate the amplitude ($A$), damping factor ($\Gamma$), frequency ($\omega$) and epoch ($\delta_0$) from which we can calculate the beating frequencies. When the amplitudes of both oscillations are the same, the beating frequency is simply the mean of two frequencies. For different amplitudes, however, the beats would not be exactly periodic and in that case we would consider the average frequency. One can also detect the topological states of either spins from the beating pattern (Fig.~\ref{topobeat}). If spin up and down belong to opposite topological states, then the oscillations are out of phase. Consequently the charge density increases initially while the spin density decreases. When the states are in the same topological phase, we observe the reverse pattern. One should note that the current is directly proportional to the time derivative of the position, and hence one can see the same oscillation in current as well. 

\begin{figurehere}
\centerline{
\includegraphics[scale=1.]{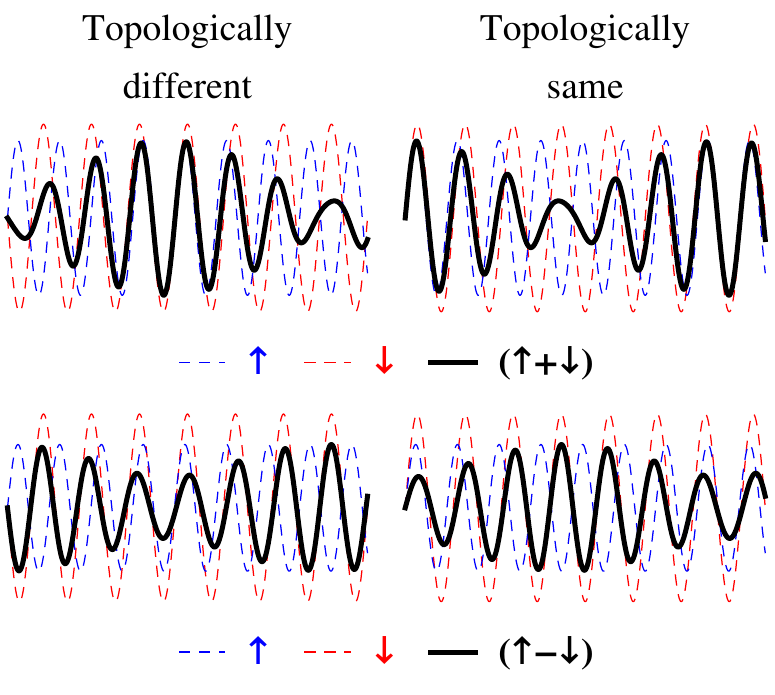}
}
\caption{Nature of beating  when spin up and down are in the same and opposite topological phases.}
\label{topobeat}
\end{figurehere}

Let us now look at the amplitude of the beating. From Fig.~\ref{topobeat} one can see that ($\uparrow-\downarrow$) oscillation does not provide any new information compared to ($\uparrow+\downarrow$) oscillations and hence we focus on ($\uparrow+\downarrow$) wavepacket only. We choose different $E_Z,M$ combination and calculate the amplitude frequency and initial phase of the resultant beating. We define a relative phase factor $\Theta = \pm1$ which indicates whether the two spin projections have the same ($\Theta = +1$) or opposite ($\Theta = -1$) initial phase, and plot the product of $\Theta$ and amplitude of the beating over the whole $E_Z-M$ space. 
Fig. \ref{xplus} displays the modulated amplitude of the longitudinal Zitterbewegung, $\Theta x_{ZB}$, when varying both $E_Z$ and $M$ for for (a) d=10000a, (b) d=5000a, and (c) d=2000a. The corresponding topological phase diagram calculated by Ezawa \cite{Ezawa2012} is indicated by the dashed lines. 
We find a good match between our analysis and the analytical phase diagram. Notice though that since the Zitterbewegung effect involves interference among states within a range of momentum, the accuracy of the boundaries between different regions of the phase diagram are sensitive to the width of the wave packet. For a spatially wide wave packet [d=10000a$\approx$300-500nm, Fig. \ref{xplus}(a)], a small number of states are involved and the boundaries are well defined. However, upon reducing the wavepacket width, more states are involved in the Zitterbewegung process and the boundaries deteriorate [d=2000a$\approx$ 60-100nm, Fig. \ref{xplus}(a)]. Therefore, a good definition of the boundaries of the topological phase diagram requires the use of a spatially wide wavepacket.

\begin{figurehere}
\centerline{
\includegraphics[scale=0.7]{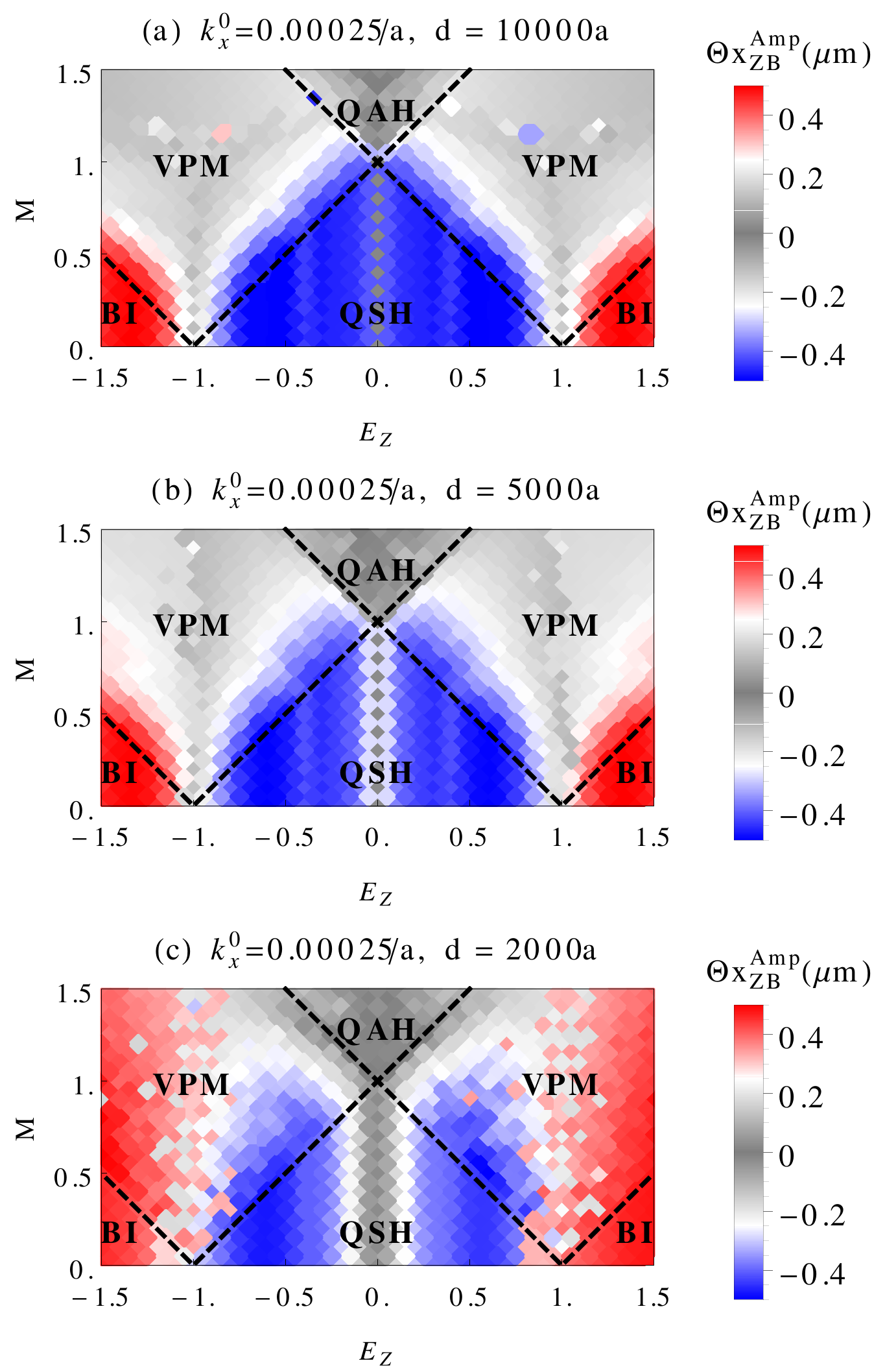}
}
\caption{Product of the beat ($\uparrow+\downarrow$) amplitude of longitudinal Zitterbewegung ($x_{ZB}^{Amp}$) and relative phase factor ($\Theta$) over the $E_Z-M$ plane for different width of momentum distribution. The top (bottom) panel corresponds to a broad (narrow) spatial distribution, i.e. a narrow (broad) momentum distribution.}
\label{xplus}
\end{figurehere}

\section{Conclusion}

We have demonstrated that Zitterbewegung features correlate with the various topological phases of two-dimensional hexagonal lattices. By analyzing the longitudinal jittering motion of the wavepacket, we were able to reconstruct the phase diagram of silicene up to a good accuracy, providing that the wavepacket considered is spatially wide. An interesting aspect of this analysis is that is provides access to the bulk properties of the material directly without the need for searching for quantized edge states. The unpolarized wavepacket described in the present work can be realized and detected using optical techniques such as pump-probe method. Such techniques have been recently exploited to investigate the ultrafast dynamics of Dirac electrons in graphene \cite{PRL157402,PRL237401,PRL257401,PRL125503}. In this context, the search for the Zitterbwegung effect and potential signatures of topological phase transition in hexagonal honeycomb lattices constitute an appealing experimental challenge.

\section*{Acknowledgement}
The research reported in this publication was supported by the King Abdullah University of Science and Technology (KAUST).

\bibliographystyle{ws-spin}
\bibliography{library}

\end{multicols}
\end{document}